\documentclass[paper,notoc]{JHEP3}
\usepackage{cite}
\usepackage{graphicx}
\usepackage{amssymb,amsfonts,amsmath,bbold,latexsym}

\title{\boldmath 
Mixed QCD-Electroweak corrections to Higgs boson production in gluon fusion
}
\author{Charalampos Anastasiou\\
  Institute for Theoretical Physics, ETH Zurich,\\
  8093 Zurich, Switzerland\\
  E-mail: \email{babis@phys.ethz.ch}}
\author{Radja Boughezal\\
Institute for Theoretical Physics,
University of Zurich,\\
Winterthurerstr. 190,
8057 Zurich, Switzerland \\ 
  E-mail: \email{radja@physik.uzh.ch}}
\author{Frank Petriello\\
 Department of Physics, University of Wisconsin,\\
 Madison, WI 53706 USA\\
  E-mail: \email{frankjp@physics.wisc.edu}}

\abstract{ 
We compute the 3-loop ${\cal O}(\alpha \alpha_s)$ correction
to the Higgs boson production cross section arising from light quarks 
using an effective theory approach.  Our calculation probes the factorization of QCD and electroweak perturbative corrections to this process.  
We combine our results with the best current estimates for contributions from top and bottom quarks to derive an 
updated theoretical prediction for the Higgs boson production cross section in gluon fusion.  With the use of the MSTW 2008 parton distribution functions that include the newest experimental data, our study results in cross sections approximately $4-6\%$ lower for intermediate Higgs boson masses than those used in recent Tevatron analyses that imposed a 95\% confidence level exclusion limit of 
a Standard Model Higgs boson with $M_H=170\,{\rm GeV}$.
} 

\begin{document}

\section{Introduction}
\label{sec:introduction}

The Higgs boson is the last undiscovered particle of the Standard Model.  The search for the Higgs is a primary goal of the LHC program, and it is also a top priority at the Tevatron.  A discovery of the Higgs boson is feasible with good confidence 
at the LHC for all mass values where the Standard Model remains consistent.  The Tevatron experiments are becoming sensitive to Higgs signals in the $H \to WW$ channel for masses near the threshold $M_H \approx 2M_W$.  Recently, the Tevatron 
collaborations reported a 95\% confidence level exclusion of a Standard Model Higgs boson with a mass near $M_H = 170 \,{\rm GeV}$~\cite{Bernardi:2008ee,Herndon:2008uv}.

Understanding the theoretical prediction is crucial to both the search for and exclusion of the Standard Model Higgs boson.  Backgrounds to the Higgs signal are severe in many channels, particularly when a mass peak cannot be reconstructed such as in 
$H\to WW \to l\nu l\nu$, and knowledge of the signal shape and normalization is needed to optimize experimental searches.  Measurements of Higgs boson couplings will also require the best possible theoretical predictions~\cite{Duhrssen:2004cv,Anastasiou:2005pd}.  
The dominant production mode at both the Tevatron and the LHC, gluon fusion through top-quark loops, receives important QCD radiative corrections~\cite{Dawson:1990zj,Djouadi:1991tka,Spira:1995rr}.  The inclusive result increases by a factor of 2 at the 
LHC and 3.5 at the Tevatron 
when perturbative QCD effects through next-to-next-to-leading order (NNLO) are taken into account~\cite{Harlander:2002wh,Anastasiou:2002yz,Ravindran:2003um}.  The theoretical uncertainty from effects beyond NNLO is estimated  to be 
about $\pm 10\%$ by varying renormalization and factorization scales.  A  better perturbative convergence and a much smaller scale variation are found when typical 
experimental cuts  which suppress associated jet radiation at high transverse momentum and 
enhance the $H \to WW$ signal at Tevatron and the LHC are implemented~\cite{Anastasiou:2004xq,Anastasiou:2005qj,Catani:2007vq,Anastasiou:2007mz,Grazzini:2008tf,Anastasiou:2008ik}. 

The importance and success in taming the QCD corrections to Higgs production have shifted attention to electroweak corrections to the Higgs signal.  The authors of Refs.~\cite{Aglietti:2004nj,Aglietti:2006yd} pointed out important 2-loop light-quark effects; these 
are pictured in Fig.~(\ref{fig:2loop}) of this manuscript and involve the Higgs coupling to $W$- or $Z$-bosons which then couple to gluons through a light-quark loop.   These terms are not suppressed by light-quark Yukawa couplings, and receive a multiplicity enhancement from summing over the quarks.  A careful study of the full 2-loop electroweak effects was performed in Ref.~\cite{Actis:2008ug}.  They increase the leading-order cross section by 
up to $5-6\%$ for relevant Higgs masses.  However, it is unclear whether these contributions receive the same large QCD enhancement as the top quark loops.  If they do, then the full NNLO QCD result is shifted by $+5-6\%$ from these electroweak corrections.  If not, this $5-6\%$ increase from light quarks would be reduced to $1-2\%$ of the NNLO result.  As this effect on the central value of the production cross section and therefore on the exclusion limits and future measurements is non-negligible, it is important to quantify it.  The exact computation of the mixed electroweak/QCD effects needed to do so requires 3-loop diagrams with 
many kinematic scales, and 2-loop diagrams with four external legs for the real-radiation terms.  Such a computation is prohibitively difficult with current computational techniques.

In this paper we compute the QCD correction to the light-quark terms in the Higgs production cross section using an effective theory approach.  We justify our approach 
rigorously by applying a hard-mass expansion procedure to the full 3-loop corrections.  This technique reduces the calculation to the evaluation of 3-loop 
vacuum bubbles.  The effective theory is formally valid only for $M_H < M_W$.  However, there are 
reasons to believe that the $K$-factor computed with the effective theory has an extended range of validity.  In the top-quark contribution to gluon fusion, the effective theory obtained after decoupling the 
top quark is formally valid only for Higgs boson masses $M_H < 2m_{t}$.  Nevertheless, the $K$-factor obtained is an extremely good approximation to the exact one for Higgs boson masses 
up to $ M_H \approx 1 \,{\rm TeV}$~\cite{Spira:1995rr}.  We find that the correction to the light-quark terms is not as large as those affecting the top-quark contribution.  Nevertheless, the two corrections have the same sign, and the numerical effect of the difference is small, indicating that the $5-6\%$ shift is indeed realized.

A second goal of this manuscript is to provide the most up-to-date QCD prediction for the Higgs boson production cross section in gluon fusion for use in setting Tevatron exclusion limits.  The CDF and D0 collaborations~\cite{cdfweb} currently use results from Ref.~\cite{Catani:2003zt}, which are several years old, augmented by the light-quark corrections from Ref.~\cite{Aglietti:2006yd}.  The following aspects of the analysis given must be updated to account for recent developments.
\begin{itemize}

\item Both CTEQ and MRST parton distribution functions (PDFs) have shifted significantly in the past several years due to an improved treatment of heavy-quark mass effects at low 
$Q^2$~\cite{Thorne:2008xf,Martin:2007bv,Nadolsky:2008zw,Martin:2009iq}, and inclusion of several Tevatron Run II data sets~\cite{Martin:2009iq}.  The 
new PDFs have a different $\alpha_s(M_Z)$ and gluon distribution, and decrease the predicted production cross section.

\item The analysis in Ref.~\cite{Catani:2003zt} used the $K$-factor computed in the effective theory with the top quark integrated out for both the top- and bottom-quark contributions.  The NLO QCD correction to the 
bottom-quark contribution is known to be smaller than the NLO top-quark $K$-factor~\cite{Spira:1995rr,Anastasiou:2006hc}.  This effect increases the predicted production cross section.

\item An updated treatment of the 2-loop light-quark contributions from Ref.~\cite{Actis:2008ug}, together with the QCD correction to these terms evaluated here, leads to a slightly smaller increase near the $M_H \approx 2M_W$ threshold than used in the Tevatron analysis.

\end{itemize}
We present results for the Higgs boson cross section accounting for these effects.  We account for the effect of soft-gluon resummation at the Tevatron by presenting values for the scale choice 
$\mu_F=\mu_R=M_H/2$, which is known to very accurately reproduce the reference value of the resummation result~\cite{Catani:2003zt} for a wide range of Higgs boson masses, and provide an 
estimate of the remaining theoretical uncertainties arising from unknown higher-order terms and PDF errors.  The updated numerical values for the cross section are approximately 
$4-6\%$ lower than those used in Tevatron analyses for Higgs boson masses in the range $150-180$ GeV, and motivate a reanalysis of the Tevatron exclusion limits.  We present a 
detailed discussion of the uncertainties arising from scale variation, PDF errors, and other theoretical effects.
  
Our paper is organized as follows.  In Section~\ref{sec:calc} we describe our calculation of the 3-loop light-quark correction to the Higgs production cross section, detailing the formulation of the effective theory and technical aspects.   In Section~\ref{sec:results} we present numerics for both the light-quark electroweak shifts and the updated inclusive cross section.  We conclude in Section~\ref{sec:conc}.

\section{Calculational Details}
\label{sec:calc}

The cross section for Higgs boson production in hadronic collisions can be written as 
\begin{eqnarray}
\sigma(s,M_H^2) &=& \sum_{i,j} \int_{0}^1 dx_1  \int_{0}^1 dx_2 \,f_{i/h_1}(x_1,\mu_F^2)  f_{j/h_2}(x_2,\mu_F^2) \int_{0}^1 dz\,\delta\left(z-\frac{M_H^2}{x_1 x_2 s}\right) \nonumber \\ &\times& z \,\hat{\sigma}_{ij}\left(z; \alpha_s(\mu_R^2), \alpha_{EW},M_H^2/\mu_R^2; M_H^2/\mu_F^2\right).
\end{eqnarray}
Here, $\sqrt{s}$ is the center-of-mass energy of the hadronic collision, $\mu_R$ and $\mu_F$ respectively denote the renormalization and factorization scales, and the $f_{i/h}$ denote the parton densities.  The quantity $z \hat{\sigma}$ is the partonic cross section for the process $ij \to H+X$ with $i,j = g,q,\bar{q}$.  As indicated, it admits a joint perturbative expansion in the strong and electroweak couplings.

The leading term in the partonic cross section arises from a one-loop correction coupling the Higgs boson to gluons via a top-quark loop.  Considering only QCD interactions for the moment and suppressing the scale dependences for simplicity, the partonic cross section can be written as 
\begin{equation}
\hat{\sigma}_{ij} = \sigma^{(0)} G_{ij}\left(z; \alpha_s \right),
\label{pcsec}
\end{equation}
with
\begin{equation}
\sigma^{(0)} = \frac{G_F \alpha_s^2}{512 \sqrt{2} \pi} \bigg| {\cal G}_t \bigg|^2,
\label{top1}
\end{equation}
\begin{equation}
{\cal G}_q = -4\,q_H\left[2-(1-4\,q_H)H\left(-r,-r; -\frac{1}{q_H}\right)\right],
\label{top2}
\end{equation}
$q_H=m_q^2/M_H^2$, and
\begin{equation}
H\left(-r,-r; x\right) = \frac{1}{2} {\rm ln}^2\left[\frac{\sqrt{4+x}-\sqrt{x}}{\sqrt{4+x}+\sqrt{x}}\right].
\end{equation}
The contribution from bottom quarks in the Standard Model is also numerically relevant; we discuss its inclusion later in this paper.  The coefficient functions can be expanded in the strong coupling constant $\alpha_s$ as
\begin{equation}
G_{ij}(z;\alpha_s) = \sum_{n=0}^{\infty} \left(\frac{\alpha_s}{\pi}\right)^n G^{(n)}_{ij}(z),
\label{cfdef}
\end{equation}
with the leading term given by $G^{(0)}_{ij}(z) = \delta_{ig}\delta_{jg} \delta(1-z)$.  The NLO coefficient functions have been computed in Ref.~\cite{Spira:1995rr} retaining the exact dependence on the quark mass.  The NNLO results in the large $m_q$ limit, 
relevant when $2\, m_q > M_H$, were 
derived in Refs.~\cite{Harlander:2002wh,Anastasiou:2002yz,Ravindran:2003um}.  The NLO result in this limit was obtained in Refs.~\cite{Dawson:1990zj,Djouadi:1991tka}. The QCD corrections have a large effect on the predicted cross section, increasing it roughly by a factor of 2 at the LHC and by a factor of 3.5 at the Tevatron.

\begin{figure}[h]
\vspace{0.3cm}
\begin{center}
\includegraphics[width=0.40\textwidth]{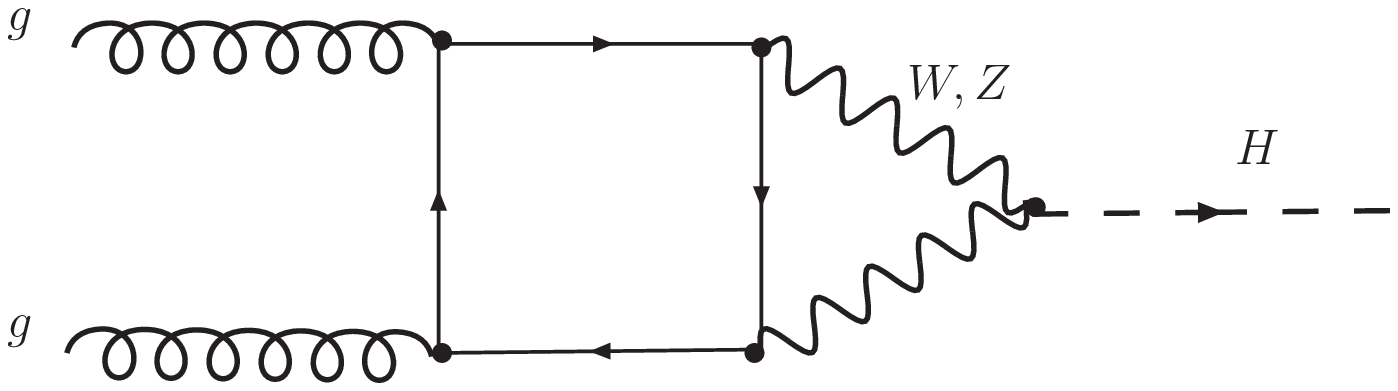}
\caption{\textsf{Example two-loop light-quark diagram contributing to the Higgs boson production cross section via gluon fusion.}}
\label{fig:2loop}
\end{center}
\end{figure}

Important electroweak corrections arise from two-loop diagrams containing an internal quark loop where the Higgs boson couples to $W$- and $Z$-bosons.  An example diagram is shown in Fig.~(\ref{fig:2loop}); we henceforth refer to these corrections as light-quark electroweak contributions, while the quark Yukawa coupling dependent terms discussed above are denoted as heavy-quark QCD contributions.  The light-quark diagrams are not suppressed by quark Yukawa couplings, and 
therefore have a multiplicity enhancement from summing over light quarks.  The inclusion of these contributions modifies the term proportional to $G^{(0)}_{ij}(z)$ in Eq.~(\ref{pcsec}).  The partonic cross section becomes
\begin{equation}
\hat{\sigma}_{ij} = \sigma^{(0)}_{\rm EW} \,G^{(0)}_{ij}\left(z\right)+\sigma^{(0)} \sum_{n=1}^{\infty} \left(\frac{\alpha_s}{\pi}\right)^n G^{(n)}_{ij}(z)
\label{pcsec2}
\end{equation}
with
\begin{equation}
\sigma^{(0)}_{\rm EW} = \frac{G_F \alpha_s^2}{512 \sqrt{2} \pi} \bigg|{\cal G}^{2l}_{lf}+{\cal G}_t \bigg|^2.
\end{equation}
${\cal G}^{2l}_{lf}$ is the expression for the two-loop light-quark contributions; its analytic form in terms of generalized harmonic polylogarithms can be found in Ref.~\cite{Aglietti:2004nj}.  A calculation 
of the corrections with the light-quark loop replaced by a top-quark, or the top-bottom doublet in the case of the $W$-boson, was first performed in Ref.~\cite{Degrassi:2004mx}.  A careful numerical study of these electroweak corrections utilizing the 
complex-mass scheme to handle the threshold regions $M_H \approx 2\, M_{W,Z}$ was performed recently in Ref.~\cite{Actis:2008ug}; this study also includes effects from internal top quarks coupling to the $W$ and $Z$.  The full corrections increase the leading-order cross section by $+5-6\%$ for Higgs boson masses in the range $120-160\,{\rm GeV}$.

The cross section in Eq.~(\ref{pcsec2}) includes corrections to the leading-order result valid through ${\cal O}(\alpha)$ in the electroweak couplings and to ${\cal O}(\alpha_s^2)$ in the QCD coupling constant in the large top-mass limit upon inclusion of the known results for $G^{(1,2)}_{ij}$.  Since the perturbative corrections to the leading-order result are large, it is important to quantify the effect of the QCD corrections on the light-quark electroweak contributions.  This would require knowledge of the mixed ${\cal O}(\alpha\alpha_s)$ 
corrections, which arise from 3-loop diagrams.  In lieu of such a calculation, the authors of Ref.~\cite{Actis:2008ug} studied two assumptions for the effect of QCD corrections on the 2-loop light-quark diagrams.
\begin{itemize}

\item {\it Partial factorization}: no QCD corrections to the light-quark electroweak diagrams are included, so that the cross section is given by the expression in Eq.~(\ref{pcsec2}).  With this assumption, electroweak diagrams contribute only a $+1-2\%$ increase to the 
Higgs boson production cross section.

\item{\it Complete factorization}: the QCD corrections to the electroweak contributions are assumed to be identical to those affecting the heavy-quark diagrams, and the partonic cross section is therefore taken to be
\begin{equation}
\hat{\sigma}^{CF}_{ij} = \sigma^{(0)}_{\rm EW} G_{ij}(z;\alpha_s)
\end{equation}
with the full QCD coefficient function multiplying both the heavy- and light-quark contributions.  In this case the light-quark diagrams increase the full NNLO QCD production cross section by $+5-6\%$.

\end{itemize}
The resulting shift in the central value for the Higgs boson production cross section can have a non-negligible effect on exclusion limits at the Tevatron, and on future exclusion limits or measurements of Higgs boson properties at the LHC.

We discuss later in this manuscript the inclusion of bottom-quark contributions to the Higgs production cross section.   We define for future reference the following point-like cross sections:
\begin{eqnarray}
\sigma^{(0)}_b &=& \frac{G_F \alpha_s^2}{512 \sqrt{2} \pi} | {\cal G}_b |^2, \nonumber \\ 
\sigma^{(0)}_{t,b} &=& \frac{G_F \alpha_s^2}{512 \sqrt{2} \pi} \left[2\,{\rm Re}\left({\cal G}_t {\cal G}_b^{*}\right)\right], \nonumber \\
\sigma^{(0)}_{t,lf} &=& \frac{G_F \alpha_s^2}{512 \sqrt{2} \pi} \left[2\,{\rm Re}\left({\cal G}_t {\cal G}_{lf}^{*}\right)\right].
\label{botcsecs}
\end{eqnarray}
$\sigma^{(0)}_b$ denotes the squared bottom-quark loop, $\sigma^{(0)}_{t,b}$ the interference between the top and bottom loops, and $\sigma^{(0)}_{t,lf}$ the interference between the top-quark contribution 
and the light-quark terms.

\subsection{The Effective Lagrangian formulation}
\label{subsec:efftheory}

A rigorous test of factorization of QCD and electroweak corrections to Higgs boson production in the Standard Model for all values of $M_H$ would require a full 3-loop calculation containing several mass scales.  Such a computation is seemingly beyond current technical capabilities.  However, in the limit $M_H < M_{W,Z}$, an approximate result can be obtained by expanding around the point $M_H/M_{W,Z} = 0$.  This is the same approach used to derive the heavy-quark result when $M_H/(2 m_t) < 1$.  
Although experimentally $M_H > M_{W,Z}$, it is known that the approximate NLO correction to the heavy-quark result matches the exact NLO correction extremely well up to $M_H \approx 1\,{\rm TeV}$ due to the structure of the QCD corrections.  This provides some reason to 
believe the same is true for the light-quark contributions.

The $M_H=0$ expansion is most clearly formulated using an effective Lagrangian, and we review this approach here.  The leading term in the expansion for the top-quark contribution in Eqs.~(\ref{top1}) and~(\ref{top2}) can be obtained via Feynman rules derived from
\begin{equation}
{\cal L}_{eff} = -\alpha_s\frac{C_1}{4v} H G_{\mu\nu}^a G^{a\mu\nu}.
\label{lag}
\end{equation}
The Wilson coefficient arising from integrating out the heavy quark is~\cite{Chetyrkin:1997iv,Chetyrkin:1997un}
\begin{eqnarray}
C_1 &=& -\frac{1}{3\pi}\left\{1+a_sC_{1q} + a_s^2 C_{2q} + {\cal O}(a_s^3) \right\}, \nonumber \\
C_{1q} &=& \frac{11}{4},\;\;\; C_{2q} = \frac{2777}{288} +\frac{19}{16}L_t+N_F\left(-\frac{67}{96}+\frac{1}{3}L_t\right),
\end{eqnarray}
where $a_s = \alpha_s/\pi$, $N_F=5$ is the number of active quark flavors and $L_t = {\rm ln}(\mu_R^2/m_t^2)$.  We now include the leading term in the $M_H/M_{W,Z}$ expansion in the Wilson coefficient.  It can be obtained by expanding ${\cal G}^{2l}_{lf}$ given 
in Ref.~\cite{Aglietti:2004nj}; denoting the contribution by $\lambda_{EW}$, we find
\begin{eqnarray}
C_1 &=& -\frac{1}{3\pi}\left\{1+\lambda_{EW}+a_sC_{1q} + a_s^2 C_{2q} + {\cal O}(a_s^3) \right\}, \nonumber \\
\lambda_{EW} &=& \frac{3\alpha}{16\pi s_W^2}\left\{\frac{2}{c_W^2}\left[\frac{5}{4}-\frac{7}{3}s_W^2+\frac{22}{9}s_W^4\right]+4\right\},
\end{eqnarray}
where $s_W,c_W$ are respectively the sine and cosine of the weak-mixing angle.

The QCD corrections modify the Wilson coefficient to include terms of ${\cal O}(\lambda_{EW}a_s)$ and  ${\cal O}(\lambda_{EW}a_s^2)$.  The extent to which factorization of electroweak and QCD corrections holds becomes a question regarding to what extent 
the Wilson coefficient can be written as a product of separate QCD and electroweak factors.  We denote the exact coefficients of these terms as $C_{1w}$ and $C_{2w}$ respectively, and introduce below in Eq.~(\ref{wilson}) the exact Wilson coefficient and the 
factorized hypothesis:
\begin{eqnarray}
C_1 &=& -\frac{1}{3\pi}\left\{1+\lambda_{EW}\left[1+a_s C_{1w}+a_s^2 C_{2w}\right]+a_sC_{1q} + a_s^2 C_{2q}  \right\}, \nonumber \\
C_1^{fac} &=& -\frac{1}{3\pi}\left(1+\lambda_{EW}\right) \left\{1+a_sC_{1q} + a_s^2 C_{2q}\right\}.
\label{wilson}
\end{eqnarray}
Factorization holds if $C_{1w}=C_{1q}$ and $C_{2w}=C_{2q}$.  We will derive here the $C_{1w}$ coefficient by expanding the 3-loop QCD corrections to the light-quark electroweak diagrams to test this.  We do not compute $C_{2w}$, but will study later 
the numerical effect of various choices for this term.

\subsection{Calculational approach}
\label{subsec:calc}

\begin{figure}
 \begin{center}
   \includegraphics[width=0.40\textwidth]{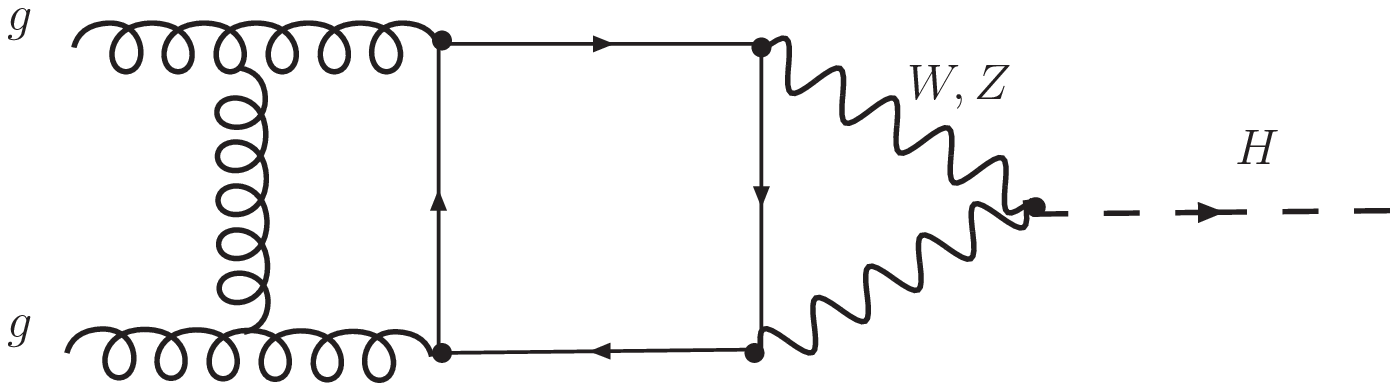}
   \includegraphics[width=0.40\textwidth]{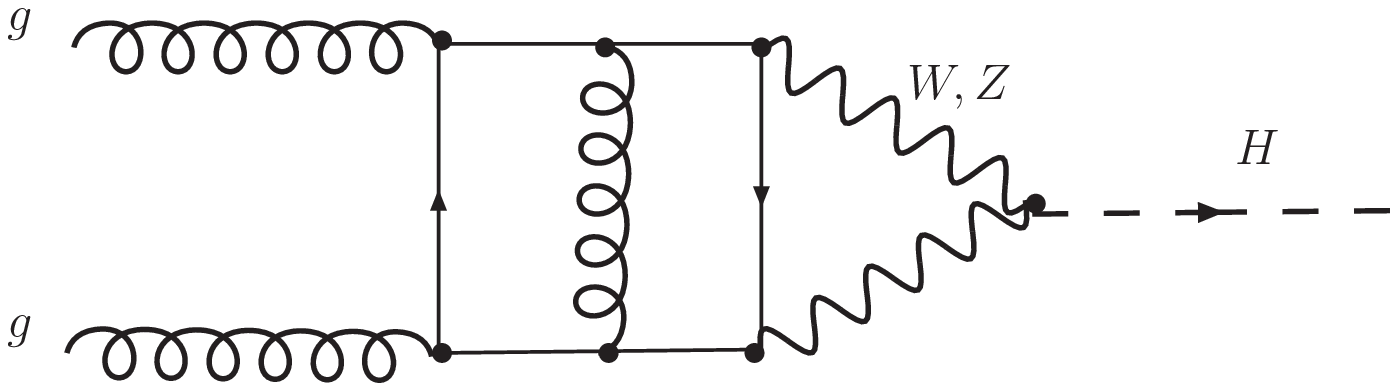}
   \caption{\textsf{Example three-loop light-quark diagrams contributing 
            to the $C_{1w}$ term in the Wilson coefficient.}}
   \label{fig:3loop}
 \end{center}
\end{figure}

We begin by generating all 3-loop diagrams for $g(p_1) + g(p_2) \to H(p_H)$ containing two internal $W$- or $Z$-boson propagators coupling to a light-quark loop.  
Examples are shown in Fig.~\ref{fig:3loop}.  Contributions 
with internal top quarks attached to the $W$ or $Z$ are very small at the 2-loop level for the Higgs boson masses we consider, as can be seen from Ref.~\cite{Actis:2008ug}, and can be 
safely neglected.  For the $Z$-boson and for a single quark flavor, there are 51 such non-vanishing diagrams.   Examples are shown in Fig.~(\ref{fig:3loop}).  
The only difference for the $W$-boson is the change of quark flavor at the vertex.  All such details are accounted for in $\lambda_{EW}$ and do not affect the computation of $C_{1w}$.  We perform a Taylor expansion of the integrand of each diagram in the external 
momenta $p_{1,2}$.  This is most conveniently performed by applying the following differential operator to each diagram~\cite{Fleischer:1994ef}:
\begin{equation}
{\cal D \,F} = \sum_{n=0}^{\infty} \left(p_1 \cdot p_2\right)^n \left[D_n {\cal F}\right]_{p_1=p_2=0}
\end{equation}
where ${\cal F}$ denotes a diagram.  The first few terms in the sum are
\begin{equation}
D_0=1,\;\;\; D_1=\frac{1}{d}\Box_{12},\;\;\; D_2=-\frac{1}{2(d-1)d(d+2)}\left\{\Box_{11}\Box_{22}-d\,\Box_{12}^2\right\},
\end{equation}
where $\Box_{ij} = \frac{\partial}{\partial p_i\mu} \frac{\partial}{\partial p_j^{\mu}}$.  The amplitude arising from summing over Feynman diagrams can be written as
\begin{equation}
\sum {\cal F} = {\cal A} \left\{g_{\mu\nu} - \frac{p_{2\mu}p_{1\nu}}{p_1 \cdot p_2}\right\} \delta^{ab}\, \epsilon^{\mu}_a (p_1) \epsilon^{\nu}_b (p_2) \equiv {\cal M}_{\mu\nu}^{ab} \epsilon^{\mu}_a (p_1) \epsilon^{\nu}_b (p_2).
\label{adef}
\end{equation} 
The coefficient ${\cal A}$ can be obtained by acting with the appropriate projection operator:
\begin{equation}
{\cal A} = \frac{1}{8\,(d-2)} \left\{g^{\mu\nu} - \frac{p_{1}^{\mu}p_{2}^{\nu}+p_{2}^{\mu}p_{1}^{\nu}}{p_1 \cdot p_2}\right\} \delta_{ab} \, {\cal M}_{\mu\nu}^{ab}.
\end{equation}
The leading term in the expansion of ${\cal A}$ must be finite, and gives $C_{1w}$ upon comparison with Eqs.~(\ref{lag}) and~(\ref{wilson}).  Sub-leading terms in the Taylor expansion need not be finite.  We note that the 2-loop light-quark contribution has a cut first at $p_H^2=M_{W,Z}^2$ because of helicity flow along the internal quark 
line~\cite{Degrassi:2004mx}, indicating that the radius of convergence of the expansion is $M_H < M_{W,Z}$.  The convergence is unchanged when the 3-loop corrections are added.

The validity of the effective theory and the Taylor expansion we utilize for $M_H < M_{W,Z}$ is most clearly seen by applying the hard-mass expansion procedure~\cite{smirnov} to the 3-loop diagrams.  This technique allows one to derive the asymptotic limit of a diagram in the limit of large internal masses via an expansion in subgraphs:
\begin{equation}
{\cal F}_{\Gamma} \sim \sum_{\gamma} {\cal F}_{\Gamma/\gamma} \circ {\cal T}_{k,p_i} {\cal F}_{\gamma}.
\end{equation}
In this expression, $\Gamma$ denotes all loop-momenta dependent pieces of the diagram ${\cal F}$.  $\gamma$ denotes the subgraphs, which are defined as those combinations 
of internal lines that contain all propagators with the heavy mass $M_{W,Z}$ and that are one-particle irreducible with respect to the massless lines.  ${\cal T}_{k,p_i}$ indicates 
the Taylor expansion of the subgraph with respect to the external momenta $p_i$ and also the loop momenta $k$ that are external to the subgraph.  ${\cal F}_{\Gamma/\gamma}$ 
is a reduced graph, which is what remains of the diagram upon removal of a subgraph.

To prove the validity of our procedure, we must show that the only subgraphs contributing to the leading $M_{W,Z}$ term are the full 3-loop diagrams themselves, and the 2-loop 
diagrams of Fig.~\ref{fig:2loop} that give $\lambda_{EW}$.  If this occurs, then all contributions are produced by the effective Lagrangian of Eq.~(\ref{lag}).  The 3-loop terms give 
$C_{1w}$, while the 1-loop reduced graphs multiplying the 2-loop subgraphs are given exactly by the first quantum corrections in the effective theory.  This is indeed what occurs.  We sketch 
briefly below the steps of the calculation.

On dimensional grounds the leading term of the coefficient ${\cal A}$ appearing in Eq.~(\ref{adef}), which comes from summing all 3-loop graphs, must scale as 
${\cal A} \sim g_{HVV} M_H^2/M_{W,Z}^2$, where $g_{HVV}$ is the $HVV$ coupling that has mass dimension one.  The subgraph obtained by expanding only the two 
massive gauge boson propagators goes as $g_{HVV}/M_{W,Z}^4$, and does not contribute to the leading term.  The possible 1-loop subgraphs contain either the Higgs boson 
coupling to two quarks, or two quarks and a gluon.  These subgraphs contribute only at $g_{HVV}/M_{W,Z}^4$ when summed.  We similarly find that the only 
2-loop subgraphs are those contained in $\lambda_{EW}$, which shows the validity of our approach.

In addition to the explicit check outlined above, we make two more remarks that indicate the validity of the effective theory.  The divergences of the effective theory 
match the universal structure of infrared divergences as given by the Catani factorization formula~\cite{Catani:1998bh}.  Also, the only other operator in the effective Lagrangian
describing Higgs interactions with gluons and massless quarks that could contribute at order $g_{HVV}/M_{W,Z}^2$ is $H\bar{q} \,\slash\!\!\!\!{D}q$~\cite{Chetyrkin:1997un}.  When inserted into a loop of light quarks and coupled to two gluons, this contribution gives scaleless integrals and vanishes; we note that this is a simple confirmation of our statement 
above that the only 1-loop reduced graphs that contribute are those generated by the effective Lagrangian of Eq.~(\ref{lag}).

We now proceed with our calculation.  All integrals appearing in ${\cal A}$ can be mapped to the following topology:
\begin{eqnarray}
{\cal I}\left(\nu_1,\nu_2,\nu_3,\nu_4,\nu_5,\nu_6\right) &=& \int d^d k_1 d^d k_2 d^d k_3 \, \frac{1}{\left[k_1^2\right]^{\nu_1} \left[k_2^2\right]^{\nu_2} \left[k_3^2-M_{W,Z}^2\right]^{\nu_3}} \nonumber \\ 
 &\times& \frac{1}{\left[(k_1-k_2)^2\right]^{\nu_4} \left[(k_2-k_3)^2\right]^{\nu_5} \left[(k_3-k_1)^2\right]^{\nu_6}}.
\end{eqnarray} 
These can be reduced to a small set of master integrals using what is by now standard technology based on the solution of integration-by-parts identities~\cite{Chetyrkin:1981qh,Laporta:2001dd,Anastasiou:2004vj}.  We find that only the integrals ${\cal I}(1,0,1,1,1,0)$ and 
${\cal I}(1,1,1,0,1,1)$ are needed to describe this process.  It is straightforward to express both as a simple product of Gamma functions.

After a computation following the approach outlined above, we obtain the primary analytic result of this paper:
\begin{equation}
C_{1w} = \frac{7}{6}.
\end{equation}
Two points should be noted regarding the comparison of this with the factorization hypothesis $C_{1w}^{fac} = C_{1q} = 11/4$.  First, there is a fairly large violation of the factorization result: $(C_{1q}-C_{1w})/C_{1w} \approx 1.4$.  However, both expressions have the same sign, and a large difference from the $+5-6\%$ shift found before does not occur.  We study in the next section the numerical effect of $C_{1w}$ and 
the unknown $C_{2w}$..

\section{Numerical Results}
\label{sec:results}

We present here numerical results for the Higgs production cross section and the shift arising from light-quark electroweak diagrams including the effect of $C_{1w}$ and $C_{2w}$.   Our purpose is two-fold: to study the numerical effect of the correction computed in the previous section, and to provide an updated prediction for the inclusive cross 
section for use in experimental studies.  We include all currently computed perturbative effects on the cross section.  These include the NNLO $K$-factor 
computed in the large-$m_t$ limit and normalized to the exact $m_t$-dependent LO result, the full light-quark electroweak correction and the ${\cal O}(\alpha_s)$ correction to this encoded in $C_{1w}$, and the bottom-quark contribution.  We define the following cross sections for use in our discussion:
\begin{eqnarray}
\sigma_{QCD}^{NNLO} &=& \sigma^{(0)}G_{ij}\left(z; \alpha_s \right) + \sigma^{(0)}_b G^{(0)}_{ij}\left(z\right) K_{bb}+\sigma^{(0)}_{t,b} G^{(0)}_{ij}\left(z\right) K_{tb}\;\;, \nonumber \\
\sigma_{EW}^{LO} &=&   \sigma^{(0)}_{t,lf}G^{(0)}_{ij} \left(z\right)\;\; , \nonumber \\
\sigma_{EW}^{NLO} &=&   \sigma^{(0)}_{t,lf} \left\{G^{(0)}_{ij}\left(z\right)\left[1+a_s(C_{1w}-C_{1q})\right] + a_sG^{(1)}_{ij}\left(z\right)\right\}\;\; , \nonumber \\
\sigma_{EW}^{NNLO} &=&  \sigma^{(0)}_{t,lf} \left\{G^{(0)}_{ij}\left(z\right)\left[1+a_s(C_{1w}-C_{1q}) +a_s^2 \left(C_{2w}-C_{2q}+C_{1q}(C_{1q}-C_{1w}\right)\right] \right.\nonumber \\
 & & +a_sG^{(1)}_{ij}\left(z\right)\left[1+a_s(C_{1w}-C_{1q})\right] + a_s^2G^{(2)}_{ij} \left. \right\} \;\; , \nonumber \\ 
\sigma_{EW}^{NNLO\; CF} &=& \sigma^{(0)}_{t,lf} G_{ij}\left(z; \alpha_s \right) \;\; , \nonumber \\ 
\sigma^{best} &=& \sigma_{QCD}^{NNLO}+\sigma_{EW}^{NNLO} \;\; .
\label{numcsecs}
\end{eqnarray}
Unless noted otherwise, all results use the MSTW 2008 distribution functions~\cite{Martin:2009iq} at the appropriate order noted in the superscript of the cross section.  
We briefly describe here the content of these several terms.  $\sigma_{QCD}^{NNLO}$ includes contributions from both top- and bottom-quark loops, with $\sigma^{(0)}_b$ 
and $\sigma^{(0)}_{t,b}$ defined in Eq.~(\ref{botcsecs}).  The QCD 
corrections to the top-quark in the large-$m_t$ limit are encoded in $G_{ij}\left(z; \alpha_s \right)$.  The NLO $K$-factors for the squared bottom-quark term $\sigma^{(0)}_b$  and the interference 
term $\sigma^{(0)}_{t,b}$ as derived from Ref.~\cite{Anastasiou:2006hc} are included in $K_{bb}$ and $K_{tb}$ respectively.  We note that both the study in Ref.~\cite{Catani:2003zt} and 
the Tevatron analysis put the bottom-quark terms in $\sigma^{(0)}$, and therefore use the same $K$-factor for both top- and bottom-quark loops.  This results in an underestimate of the cross section, 
since the effect of these terms is negative; while the NNLO $K-$factor for the top-quark term is roughly 2.1 and the NLO $K$-factor is roughly 1.8 with MSTW2008 PDFs, $K_{bb}$ and $K_{tb}$ only vary between 1.2 and 1.5 for Higgs boson masses between $120-180$ GeV. 

The remaining terms in Eq.~(\ref{numcsecs}) denote the contributions from light-quark diagrams in various approximations.  $\sigma^{(0)}_{t,lf}$ denotes the interference between the dominant top-quark term 
and the light quarks defined in Eq.~(\ref{botcsecs}); in our numerics we use the exact results of Ref.~\cite{Actis:2008ug} which are valid for arbitrary Higgs boson masses.  The squared light-quark term is numerically irrelevant.  $\sigma_{EW}^{LO}$ includes only the 2-loop diagrams computed in 
Refs.~\cite{Aglietti:2004nj,Actis:2008ug} and is equivalent to the partial factorization hypothesis defined in Sec.~\ref{sec:calc}.  $\sigma_{EW}^{NLO}$ includes the ${\cal O}(\alpha_s)$ correction to these 
diagrams computed in the effective theory and encoded in $C_{1w}$.  $\sigma_{EW}^{NNLO}$ includes the full ${\cal O}(\alpha_s^2)$ correction to the light-quark diagrams including the unknown 
coefficient $C_{2w}$.  We study numerically below various choices for $C_{2w}$.  $\sigma_{EW}^{NNLO\; CF}$ is the complete factorization 
hypothesis defined in Sec.~\ref{sec:calc}.  Finally, $\sigma^{best}$ is the current best prediction for the Higgs boson cross section including all effects of top, bottom, and light quarks with the best 
estimates of their associated QCD corrections.  To approximately implement the effects of soft-resummation, we make the scale choice $\mu_R=\mu_F=\mu=M_H/2$.  This choice is known to 
reproduce the central value of the resummation results to better than 1\% accuracy~\cite{Catani:2003zt}.  It has been pointed out that
the choice $\mu=M_H$ may not correctly describe the Higgs production process~\cite{Anastasiou:2002yz,Anastasiou:2005qj}, and that the perturbative convergence is improved
for $\mu=m_H/2$~\cite{Moch:2005ky}.  The effect of resummation is also smaller for this scale choice~\cite{Ahrens:2008nc}.  We evaluate the electroweak corrections using $G_F$, $M_W$ and $M_Z$ as input parameters.  We use the pole mass $m_t = 170.9\;{\rm GeV}$ for the top quark and the $\overline{MS}$ mass $\bar{m}_b(\mu_R)$ for the $b$-quark with the input value
 $\bar{m}_b(10\;{\rm GeV}) = 3.609 \;{\rm GeV}$~\cite{Kuhn:2007tn} .  The choice of pole or $\overline{MS}$ mass for the top quark has a negligible effect on the numerical results. 

We comment here on the numerical validity of the large-$m_t$ limit for the top-quark squared contribution.  Factoring out the exact top-quark dependence and computing the 
$K$-factors in the effective theory, as we do here, gives an exceptionally accurate approximation to the full result.  We have checked with an exact
calculation at NLO that our result agrees to better than 1\% in the kinematic range relevant for the Tevatron studies; confirmations of this result have been obtained using 
several independent codes~\cite{ztalks}.  Finite $m_t$ corrections to the NNLO coefficient function have been 
shown to affect the $K$-factor by less than 1\% when the full $m_t$ dependence is factored out~\cite{Schreck:2007um,Marzani:2008ih}.  We note 
that we compute exactly at NLO the top-bottom interference and bottom-bottom diagrams.  We conclude that the error arising from our treatment of the top quark mass is at 
the percent level or less.

\begin{figure}
 \begin{center}
   \includegraphics[width=0.49\textwidth]{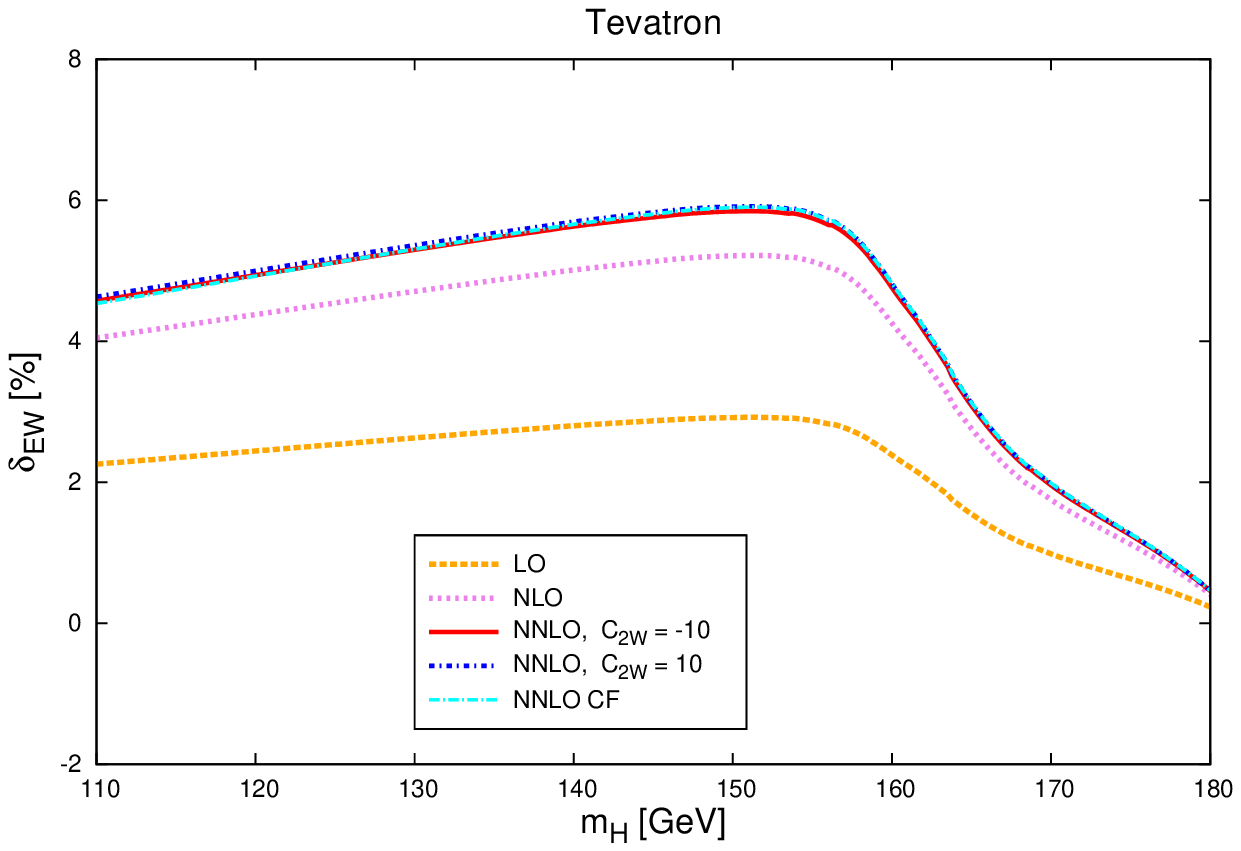}
   \includegraphics[width=0.49\textwidth]{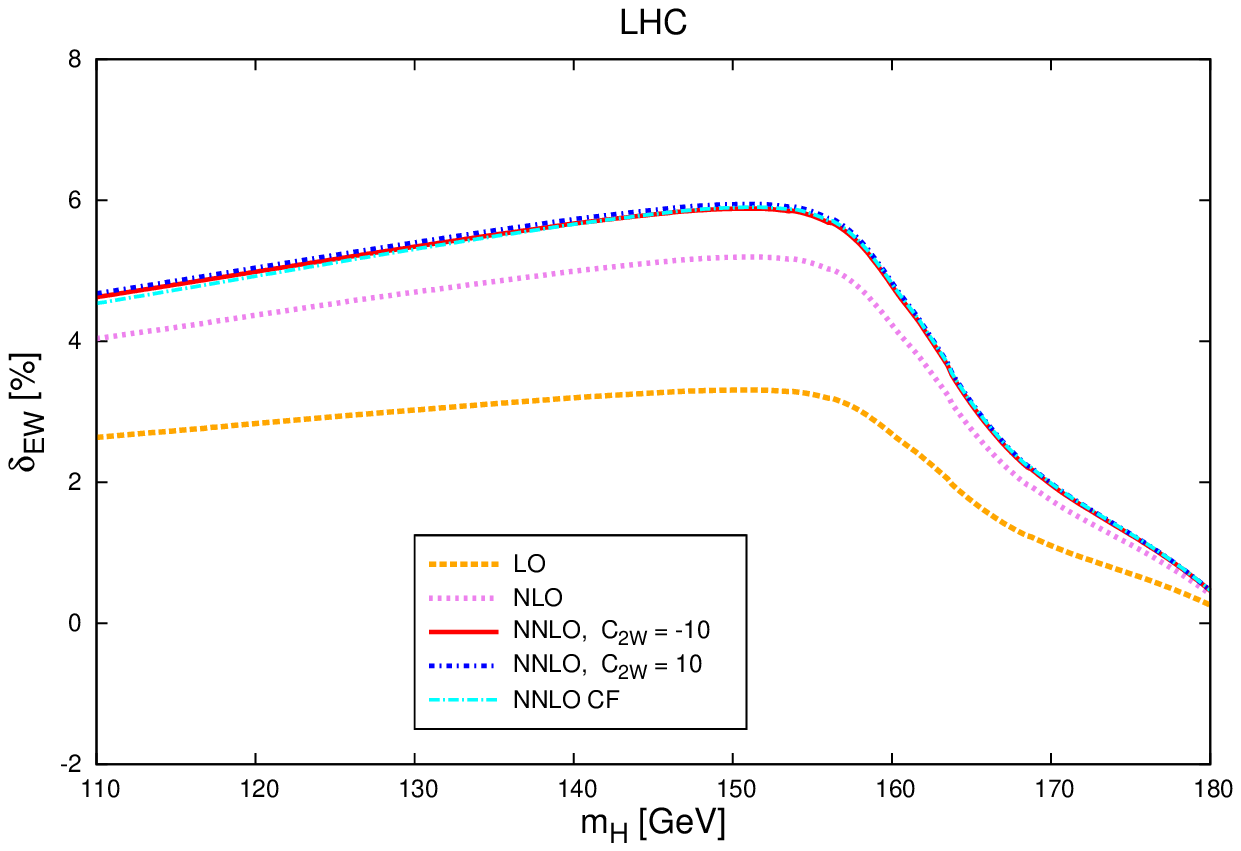}
 \caption{\textsf{Relative shifts to the Higgs boson production cross section at
    the Tevatron (left panel) and LHC (right panel) arising from light-quark diagrams.   All curves are normalized to the full NNLO top-quark 
    cross section and are produced for renormalization and factorization
    scales $\mu_R=\mu_F=M_H/2$.  
    The various lines are described in detail in the accompanying text.}}
    \label{fig:TEVshifts}
 \end{center}
\end{figure}

We begin by studying the percentage shifts arising from electroweak effects on the Higgs boson production cross section at both the Tevatron and the LHC in Fig.~(\ref{fig:TEVshifts}).  The results shown in these plots are $\delta_{EW}^x = 100 \times \sigma_{EW}^x/\sigma_{QCD}^{NNLO}$, with the cross sections defined in Eq.~(\ref{numcsecs}).  The close agreement between $\delta_{EW}^{NNLO}$ and 
$\delta_{EW}^{NNLO\,CF}$ occurs 
because the differences $C_{1w}-C_{1q}$ and $C_{2w}-C_{2q}$ in Eq.~(\ref{numcsecs}) are small compared to the effects of $G^{(1,2)}_{ij}$ in $\sigma_{EW}^{NNLO}$.  Furthermore, the unknown $C_{2w}$ 
coefficient does not significantly alter the size of the electroweak shift.  Our calculation confirms that the Higgs boson production cross section receives almost the entire numerical correction indicated by 
the complete factorization hypothesis.

We now combine all effects into a best prediction for the Higgs boson production cross section, $\sigma^{best}$ defined in Eq.~(\ref{numcsecs}). We set $C_{2w}=0$ to produce these numbers.  As discussed previously, several updates must be performed 
to the analysis in Ref.~\cite{Catani:2003zt} and therefore the Tevatron exclusion limits.
\begin{itemize}

\item The $K$-factors for $\sigma^{(0)}_b$ and $\sigma^{(0)}_{t,b}$ are now known to be significantly smaller than those for top-quark term $\sigma^{(0)}$~\cite{Spira:1995rr,Anastasiou:2006hc}.  For 
example, in the pole-mass scheme the NLO $K$-factor for the top-bottom interference is approximately 1.3 for $M_H=150\;{\rm GeV}$, as compared to over 1.7 at NLO and 3 at NNLO for the top-quark term; the $K$-factor for the bottom-quark piece is roughly 1.5 for this Higgs mass.  The study in Ref.~\cite{Catani:2003zt} utilized the top-quark $K$-factor for all three terms.  As mentioned above we use the $\overline{MS}$ $b$-quark mass instead of the pole mass of Ref.~\cite{Catani:2003zt}.  
Our final cross section numbers with the $\overline{MS}$ $b$-mass are about 1.5\% larger than the numbers computed in the pole scheme.  Although the results in both schemes are 
very similar after the inclusion of NLO corrections, previous studies have shown some preference for a running mass~\cite{Braaten:1980yq,Spira:1995rr} so we present numbers for this scheme.

\item Updated PDF extractions by CTEQ and MRST with an improved treatment of heavy-quark effects at low $Q^2$ have a significant effect on the cross section.

\item The electroweak terms derived above must be added.

\end{itemize}
These corrections all have an important effect on the predicted cross section.  For illustration, we show below the sequential effect of making these changes on the cross section for $M_H=170\;{\rm GeV}$.  
We begin by reproducing the $\mu_F=\mu_R=M_H/2$ numbers given in Ref.~\cite{Catani:2003zt}, which also matchs the reference value for the 
resummed result, by implementing MRST 2002 NNLO PDFs, using the masses $m_t=176 \;{\rm GeV}$ and $m_b=4.75\;{\rm GeV}$, multiplying all top- and bottom-quark terms by the $K$-factor 
appropriate for the top quark, and removing the 2-loop light-quark terms.  We then perform the following changes: (1) we switch to MRST 2008 NNLO PDFs; (2) we switch to the current 
extracted top-quark mass $m_t=170.9\;{\rm GeV}$, and to $\bar{m}_b$, and use the NLO $K_{tb}$ and $K_{bb}$ respectively to model 
the QCD corrections to the top-bottom interference and the bottom-quark squared contribution; (3) we implement the electroweak corrections described above.  The results are shown in   
Table~\ref{tab:mh170}.  The effect of moving to MSTW 2008 PDFs is large and negative, although the other effects compensate to a large degree.   We apply all of these corrections to provide up-to-date values for the inclusive Higgs boson production cross section in Table~\ref{tab:TEVxsec08} for the scale choice $\mu=M_H/2$ that accurately reproduces the effect of soft-gluon resummation.  These numerical values are $4-6\%$ lower than 
values previously used by the Tevatron collaborations to establish exclusion limits on the Standard Model Higgs boson~\cite{cdfweb} for $M_H=150-170$ GeV, and motivate a reconsideration of their results.  The cross section for the exclusion point is reduced by 6\% from what was used in the Tevatron analysis.
\begin{table}[t]
  \begin{center}
    \begin{tabular}{|c|c|c|c|}
      \hline
         original & MSTW 2008 PDFs & $K_{tb}$, $K_{bb}$ & EW effects \\ \hline \hline
         0.3542 & 0.3212 & 0.3377 & 0.3444 \\\hline
          \end{tabular}
  \end{center}
  \caption{\textsf{Shifts in the Higgs boson production cross section resulting from the changes detailed in the text.  All numbers are in picobarns.}
    \label{tab:mh170} }
\end{table}

We also estimate the current theoretical uncertainties arising from uncalculated higher-order terms and PDF errors.  To estimate the errors from higher-order effects we vary the scale $\mu$ in the 
range $[M_H/4,M_H]$, which is a factor of two around the central value $\mu_R=\mu_F=\mu=M_H/2$.  For the PDF errors we use the error eigenvectors provided with the MRST 2008 fit.  The scale 
errors are constant with Higgs mass to very good approximation, and are $[-11\%,+7\%]$.  The PDF uncertainties have a slight dependence on the Higgs boson mass, as shown in 
Table~\ref{tab:TEVxsec08}.
\begin{table}[t]
  \begin{center}
    \begin{tabular}{|c|c||c|c|}
      \hline
      $m_{H}$[GeV] &$\sigma^{best}$[pb]&$m_{H}$[GeV] &$\sigma^{best}$[pb]\\
        \hline \hline
     110&    1.417 ($\pm 7\%$ pdf) &160&0.4344 ($\pm 9\%$ pdf)
 \\[1mm]
      \hline
  115&    1.243 ($\pm 7\%$ pdf) &165&0.3854 ($\pm 9\%$ pdf)
\\[1mm]
      \hline
  120&    1.094  ($\pm 7\%$ pdf)  &170&0.3444 ($\pm 10\%$ pdf)
\\[1mm]
      \hline
  125 &   0.9669 ($\pm 7\%$ pdf) &175&0.3097 ($\pm 10\%$ pdf)
\\[1mm]
      \hline
  130 &  0.8570 ($\pm 8\%$ pdf) &180&0.2788 ($\pm 10\%$ pdf)
\\[1mm]
      \hline
  135 &   0.7620 ($\pm 8\%$ pdf) &185&0.2510 ($\pm 10\%$ pdf)
\\[1mm]
      \hline
  140 &   0.6794 ($\pm 8\%$ pdf) &190&0.2266 ($\pm 11\%$ pdf)
\\[1mm]
      \hline
  145 &   0.6073 ($\pm 8\%$ pdf) &195&0.2057 ($\pm 11\%$ pdf)
\\[1mm]
      \hline
  150 &   0.5439 ($\pm 9\%$ pdf) &200&0.1874 ($\pm 11\%$ pdf)
  \\[1mm]
      \hline
  155 & 0.4876  ($\pm 9\%$ pdf) & $-$& $-$
      \\\hline
    \end{tabular}
  \end{center}
  \caption{\textsf{Higgs production cross
               section (MSTW08) for Higgs mass values relevant for Tevatron, with
               $\mu= \mu_R =\mu_F=M_H/2 $. The total cross section
               $\sigma^{best}$ is defined in Eq.~(\ref{numcsecs}).  The theoretical errors PDFs are shown in the Table; the scale variation is $^{+7\%}_{-11\%}$, roughly constant as a 
               function of Higgs boson mass.  Other potential sources of theoretical error are discussed in the text.}
    \label{tab:TEVxsec08} }
\end{table}

\section{Conclusions}
\label{sec:conc}

In this paper, we considered mixed QCD-electroweak corrections to the Higgs boson 
production cross section in the gluon-fusion channel.
Working in an effective field theory valid for $m_{H}< M_{W}$,
we provided the leading term of the three-loop 
$\mathcal{O}(\alpha\alpha_s)$ contributions due to diagrams containing 
light quarks.  This result allows us to check the factorization of electroweak and QCD 
corrections proposed in Ref.~\cite{Aglietti:2006yd,Actis:2008ug}.  We showed that, despite a fairly
large violation of the factorization hypothesis, a significant numerical difference from
the prediction of this hypothesis is not observed due to the structure of the QCD corrections.  We combined the 2-loop 
light-quark diagrams based on 
the complex-mass scheme for the $W$- and $Z$-bosons~\cite{Actis:2008ug} with our new 3-loop $\mathcal{O}(\alpha\alpha_s)$ correction and 
with contributions from top and bottom quarks to provide an updated theoretical prediction for the production cross section.  
We found values that are $4-6\%$ lower than those currently used by the Tevatron collaborations in the analysis that 
led to the 95\% confidence level exclusion on a Standard Model Higgs boson with $M_H = 170\;{\rm GeV}$.  
Our results motivate a reconsideration of the Tevatron exclusion limits.

\section*{Acknowledgments}

We thank the authors of Ref.~\cite{Actis:2008ug}
for providing us with a table of the electroweak corrections
based on their 2-loop result in the complex mass scheme.
We also thank the authors of Ref.~\cite{Aglietti:2006yd} for 
providing us with Fortran routines for the numerical integration
of the GHPLs that appear in their paper based on the real mass
scheme.  We thank S.~Bucherer for numerical results for the NLO corrections to the bottom-quark contributions, and F.~Stoeckli for discussions regarding the importance of these effects.   Useful discussions with S.~Actis, T.~Becher, R.~Bonciani, T.~Gehrmann, M.~Herndon, Z.~Kunszt and I.~Rothstein are also acknowledged.  The work of C.~A. and R.~B. is supported by the Swiss National Science Foundation under contracts 200021-117873 and 200020-116756/2.  
The work of F.~P. is supported by the DOE grant DE-FG02-95ER40896, Outstanding  Junior Investigator Award and
by the Alfred P.~Sloan Foundation.

\smallskip
\noindent
{\bf Note added}: After this manuscript was first submitted, the MSTW collaboration released an update of their PDF distribution which includes new Tevatron Run II data, and is meant to supersede their previous releases~\cite{Martin:2009iq}.  The new distribution has a lower value of $\alpha_s(M_Z)$ and uses the Tevatron Run II dijet data.  The effect on the Higgs cross section of changing from MRST 2006 to MSTW 2008 
is large; the cross section shifts downwards by nearly 15\% from MRST 2006.  For illustrative purposes, we include 
below in Table~\ref{tab:pdfs} the effect on the total Higgs cross section for $M_H=170$ GeV coming from 
the change in PDF sets from 2001 to 2008.  The PDF uncertainties as estimated by the error eigenvectors have 
also increased significantly in the MSTW 2008 distribution.
   
\begin{table}[htbp]
  \begin{center}
    \begin{tabular}{|c|c|c|c|}
      \hline
         MRST 2001 & MRST 2004 & MRST 2006 & MSTW 2008 \\ \hline \hline
          0.3833 & 0.3988 & 0.3943 & 0.3444 \\\hline
          \end{tabular}
  \end{center}
  \caption{\textsf{The Higgs production cross section in picobarns for $M_H=170$ GeV at the Tevatron, using several different PDF distributions.}
    \label{tab:pdfs} }
\end{table}


\begin{thebibliography}{99}

\bibitem{Bernardi:2008ee}
  G.~Bernardi {\it et al.}  [Tevatron New Phenomena Higgs Working Group and
                  CDF Collaboration and D],
  arXiv:0808.0534 [hep-ex].

\bibitem{Herndon:2008uv}
  M.~Herndon, for the Babar, CDF and D0 collaborations,
  arXiv:0810.3705 [hep-ex].

\bibitem{Duhrssen:2004cv}
  M.~Duhrssen, S.~Heinemeyer, H.~Logan, D.~Rainwater, G.~Weiglein and D.~Zeppenfeld,
  Phys.\ Rev.\  D {\bf 70}, 113009 (2004)
  [arXiv:hep-ph/0406323].

\bibitem{Anastasiou:2005pd}
  C.~Anastasiou, K.~Melnikov and F.~Petriello,
  Phys.\ Rev.\  D {\bf 72}, 097302 (2005)
  [arXiv:hep-ph/0509014].






\bibitem{Spira:1995rr}
D.~Graudenz, M.~Spira and P.~M.~Zerwas,
  Phys.\ Rev.\ Lett.\  {\bf 70}, 1372 (1993);
  M.~Spira, A.~Djouadi, D.~Graudenz and P.~M.~Zerwas,
  Nucl.\ Phys.\  B {\bf 453}, 17 (1995)
  [arXiv:hep-ph/9504378].


\bibitem{Dawson:1990zj}
  S.~Dawson,
  Nucl.\ Phys.\  B {\bf 359}, 283 (1991).
\bibitem{Djouadi:1991tka}
  A.~Djouadi, M.~Spira and P.~M.~Zerwas,
  Phys.\ Lett.\  B {\bf 264}, 440 (1991).


\bibitem{Harlander:2002wh}
  R.~V.~Harlander and W.~B.~Kilgore,
  Phys.\ Rev.\ Lett.\  {\bf 88}, 201801 (2002)
  [arXiv:hep-ph/0201206].

\bibitem{Anastasiou:2002yz}
  C.~Anastasiou and K.~Melnikov,
  Nucl.\ Phys.\  B {\bf 646}, 220 (2002)
  [arXiv:hep-ph/0207004].
\bibitem{Ravindran:2003um}
  V.~Ravindran, J.~Smith and W.~L.~van Neerven,
  Nucl.\ Phys.\  B {\bf 665}, 325 (2003)
  [arXiv:hep-ph/0302135].

\bibitem{Anastasiou:2004xq}
  C.~Anastasiou, K.~Melnikov and F.~Petriello,
  Phys.\ Rev.\ Lett.\  {\bf 93}, 262002 (2004)
  [arXiv:hep-ph/0409088].
\bibitem{Anastasiou:2005qj}
  C.~Anastasiou, K.~Melnikov and F.~Petriello,
  Nucl.\ Phys.\  B {\bf 724}, 197 (2005)
  [arXiv:hep-ph/0501130].


\bibitem{Catani:2007vq}
  S.~Catani and M.~Grazzini,
  Phys.\ Rev.\ Lett.\  {\bf 98}, 222002 (2007)
  [arXiv:hep-ph/0703012].
\bibitem{Anastasiou:2007mz}
  C.~Anastasiou, G.~Dissertori and F.~Stockli,
  JHEP {\bf 0709}, 018 (2007)
  [arXiv:0707.2373 [hep-ph]].
\bibitem{Grazzini:2008tf}
  M.~Grazzini,
  JHEP {\bf 0802}, 043 (2008)
  [arXiv:0801.3232 [hep-ph]].
\bibitem{Anastasiou:2008ik}
  C.~Anastasiou, G.~Dissertori, F.~Stockli and B.~R.~Webber,
  JHEP {\bf 0803}, 017 (2008)
  [arXiv:0801.2682 [hep-ph]].


\bibitem{Aglietti:2004nj}
  U.~Aglietti, R.~Bonciani, G.~Degrassi and A.~Vicini,
  Phys.\ Lett.\  B {\bf 595}, 432 (2004)
  [arXiv:hep-ph/0404071].


\bibitem{Aglietti:2006yd}
  U.~Aglietti, R.~Bonciani, G.~Degrassi and A.~Vicini,
  arXiv:hep-ph/0610033.

\bibitem{Actis:2008ug}
  S.~Actis, G.~Passarino, C.~Sturm and S.~Uccirati,
  arXiv:0809.1301 [hep-ph];
S.~Actis, G.~Passarino, C.~Sturm and S.~Uccirati,
  arXiv:0809.3667 [hep-ph].


\bibitem{cdfweb}
For a discussion of the Higgs boson cross sections used by the Tevatron collaborations, see
\verb+http://www-cdf.fnal.gov/physics/new/hdg/results/hwwmenn_080725/+ .

\bibitem{Catani:2003zt}
  S.~Catani, D.~de Florian, M.~Grazzini and P.~Nason,
  JHEP {\bf 0307}, 028 (2003)
  [arXiv:hep-ph/0306211].

\bibitem{Martin:2007bv}
  A.~D.~Martin, W.~J.~Stirling, R.~S.~Thorne and G.~Watt,
  Phys.\ Lett.\  B {\bf 652}, 292 (2007)
  [arXiv:0706.0459 [hep-ph]].
  
\bibitem{Nadolsky:2008zw}
  P.~M.~Nadolsky {\it et al.},
  Phys.\ Rev.\  D {\bf 78}, 013004 (2008)
  [arXiv:0802.0007 [hep-ph]].

\bibitem{Martin:2009iq}
  A.~D.~Martin, W.~J.~Stirling, R.~S.~Thorne and G.~Watt,
  arXiv:0901.0002 [hep-ph].

\bibitem{Thorne:2008xf}
  For a review of this issue, see R.~S.~Thorne and W.~K.~Tung,
  arXiv:0809.0714 [hep-ph].

\bibitem{Anastasiou:2006hc}
  C.~Anastasiou, S.~Beerli, S.~Bucherer, A.~Daleo and Z.~Kunszt,
  JHEP {\bf 0701}, 082 (2007) and work in progress.



\bibitem{Degrassi:2004mx}
  G.~Degrassi and F.~Maltoni,
  Phys.\ Lett.\  B {\bf 600}, 255 (2004)
  [arXiv:hep-ph/0407249].

\bibitem{Chetyrkin:1997iv}
 K.~G.~Chetyrkin, B.~A.~Kniehl and M.~Steinhauser,
 Phys.\ Rev.\ Lett.\  {\bf 79} (1997) 353
 [arXiv:hep-ph/9705240].

\bibitem{Chetyrkin:1997un}
  K.~G.~Chetyrkin, B.~A.~Kniehl and M.~Steinhauser,
  Nucl.\ Phys.\  B {\bf 510}, 61 (1998)
  [arXiv:hep-ph/9708255].

\bibitem{Fleischer:1994ef}
  J.~Fleischer and O.~V.~Tarasov,
  Z.\ Phys.\  C {\bf 64}, 413 (1994)
  [arXiv:hep-ph/9403230].

\bibitem{smirnov}
{\it Applied Asymptotic Expansions in Momenta and Masses,} V.A. Smirnov, Springer Tracts in Modern Physics (2002).

\bibitem{Catani:1998bh}
  S.~Catani,
  Phys.\ Lett.\  B {\bf 427}, 161 (1998)
  [arXiv:hep-ph/9802439].

\bibitem{Chetyrkin:1981qh}
  K.~G.~Chetyrkin and F.~V.~Tkachov,
  Nucl.\ Phys.\  B {\bf 192}, 159 (1981).
  
\bibitem{Laporta:2001dd}
  S.~Laporta,
  Int.\ J.\ Mod.\ Phys.\  A {\bf 15}, 5087 (2000)
  [arXiv:hep-ph/0102033].

\bibitem{Anastasiou:2004vj}
  C.~Anastasiou and A.~Lazopoulos,
  JHEP {\bf 0407}, 046 (2004)
  [arXiv:hep-ph/0404258].

\bibitem{Moch:2005ky}
  S.~Moch and A.~Vogt,
  Phys.\ Lett.\  B {\bf 631}, 48 (2005)
  [arXiv:hep-ph/0508265].

\bibitem{Ahrens:2008nc}
  V.~Ahrens, T.~Becher, M.~Neubert and L.~L.~Yang,
  arXiv:0809.4283 [hep-ph].

\bibitem{Kuhn:2007tn}
  J.~H.~Kuhn, M.~Steinhauser and C.~Sturm,
  arXiv:0705.2335 [hep-ph].

\bibitem{ztalks}
See the talks by G. Degrassi and R. Harlander at the Workshop on Higgs Boson Phenomenology, 7-9 January 2009, Zurich, Switzerland, at the 
site \verb+http://www.itp.uzh.ch/events/higgsboson2009/index.html+.

\bibitem{Schreck:2007um}
  M.~Schreck and M.~Steinhauser,
  Phys.\ Lett.\  B {\bf 655}, 148 (2007)
  [arXiv:0708.0916 [hep-ph]].

\bibitem{Marzani:2008ih}
  S.~Marzani, R.~D.~Ball, V.~Del Duca, S.~Forte and A.~Vicini,
  arXiv:0809.4934 [hep-ph].

\bibitem{Braaten:1980yq}
  E.~Braaten and J.~P.~Leveille,
  Phys.\ Rev.\  D {\bf 22}, 715 (1980).


\end{thebibliography}
\end{document}